\documentclass{aa}
\usepackage{graphicx}

\begin{document}

\title{On the Horizontal Branch of the Galactic Globular NGC2808}

\author{V. Castellani\inst{1,2}, G. Iannicola\inst{1}, G. Bono\inst{1}, 
M. Zoccali\inst{3}, S. Cassisi\inst{4}, R. Buonanno\inst{5}}

\offprints {V. Castellani, \email {vittorio@mporzio.astro.it}}

\institute{1- INAF-Osservatorio Astronomico di Roma, via Frascati 33, 
00040 Monte Porzio Catone, Italy.\\ 
2- INFN-Sezione di Ferrara, via Paradiso 12, 44100 Ferrara, Italy.\\ 
3- Universidad Catolica de Chile, Department of Astronomy \& Astrophysics, 
Casilla 306, Santiago 22, Chile.\\   
4- INAF-Osservatorio Astronomico di Teramo, via M. Maggini, 64100 Teramo, Italy.\\  
5- Dipartimento di Fisica, Universit\`a di Roma Tor Vergata, via della Ricerca 
Scientifica 1, 00133 Rome, Italy.}

\date{Received ; accepted }

\authorrunning{Castellani et al.}

\titlerunning{HB stars in NGC2808}

\abstract{
We present new UV (F218W) data for stars in  the central region of the
Galactic globular cluster NGC2808,  collected with the WFPC2 camera on
board of the  Hubble Space Telescope.  These data together  with F439W
and F555W-band data and  previous ground based observations  provide a
multifrequency coverage of the cluster stellar population extending up
to a distance of  1.7 times the  cluster core radius. We  discuss this
complete sample  of stars, which includes  764  Red Giant Branch (RGB)
stars brighter than the Horizontal Branch  (HB) luminosity level, 1239
HB  stars,  119 Asymptotic  Giant  Branch (AGB),  and 22  AGB-manqu\'e
stellar  structures.   As already known, we   find that  blue HB stars
separate into three distinct groups. However, our multiband photometry
indicates that several  stars in the two  hotter HB groups show a flat
spectrum, thus suggesting the  binarity of these objects.   Artificial
star  experiments  suggest  that at    most  50\% of   them might   be
photometric  blends.  Moreover, at   variance with previous claims one
finds that canonical Zero Age Horizontal Branch (ZAHB) models do reach
effective temperatures typical of observed hot HB  stars. We also show
that    the ratio between   HB  and RGB  stars brighter    than the HB
luminosity level steadly increases when moving from the cluster center
to the periphery, passing from R=$1.37\pm0.14$  in the cluster core to
R=$1.95\pm0.26$ in the outer cluster regions.  We discuss the possible
origin of  such a  radial gradient in  the context  of the Blue  Tails
phenomenon, advancing    some suggestions concerning  the  clumpy
stellar distribution along the HB.

}

\maketitle

\section{Introduction}

For about   thirty years  NGC2808 has been   a  special conundrum  for
stellar  evolution understanding    in   Galactic   Globular  Clusters
(GGCs).  It was indeed the    year  1974 when  Harris disclosed    the
unexpected bimodal distribution  of HB  stars in NGC2808. 
He found that HB stars populate two well separated regions of
the  branch, either to    the red or  to  the  blue  of the   RR Lyrae
instability strip.  At  that time, it was  already clear  that such an
odd  occurrence  could be easily  understood   in terms  of a  bimodal
distribution  of  HB masses  and,  thus,  of  mass    loss in  the  HB
progenitors, the origin     of such bimodality remaining,     however,
unknown.
Owing to its peculiar HB morphology, among the known GGCs, it was 
dubbed {\em anomalous} by Harris (1978). 

This empirical  scenario became  even more  puzzling a few  years ago,
when Sosin et  al.  (1997) took  advantage of data collected  with the
Wide Field Planetary Camera  2 (WFPC2)  on  board of the  Hubble Space
Telescope (HST) to provide a deep color magnitude diagram (CMD) of the
cluster. They found  in the hot  portion of  the  HB a  long blue tail
extending over  more than 4 V-magnitudes,   with two additional narrow
but  well-defined gaps.  More recently, Sohn  et al. (1998) detected in
this  cluster a significant   color gradient, the  mean color becoming
redder when moving toward the cluster center. This would be a relevant
evidence,  possibly  connected  with   the  origin  of  the   hot   HB
tail. However, further  photometric investigations (Walker 1999; Bedin
et  al. 2000), failed  to find any   firm evidence of radial gradients
across   the cluster in  the  different evolutionary phases across the
cluster.

To investigate  such an  issue in more details,  and, in general,  to
reach a  better knowledge of the  cluster HB population, in this paper
we make use of the F218W filter of the WFPC2 on board of HST to secure
a robust detection  of the faintest blue  HB  objects in the  crowded
cluster  central    region. On  this   basis,   we  will  discuss  the
evolutionary status  of hot  HB  stars,  presenting evidence  for  the
occurrence in  the  cluster of   a population gradient.  Even  though
population     gradients represented  the    original   goal  of   our
investigation, we found that  multiband photometric data are  of great
relevance  to investigate the  nature  of hot  HB structures. In fact,
present data suggest that a   substantial fraction of subluminous   HB
stars might be binary systems. Taking into account this new empirical
evidence,   in the final  section of  this paper   we will revisit the
problem  of   multimodal HB  distributions,  suggesting  some possible
evolutionary scenarios.


\section {Observations and data reduction}

\begin{figure*}
\centering
\includegraphics[width=15cm]{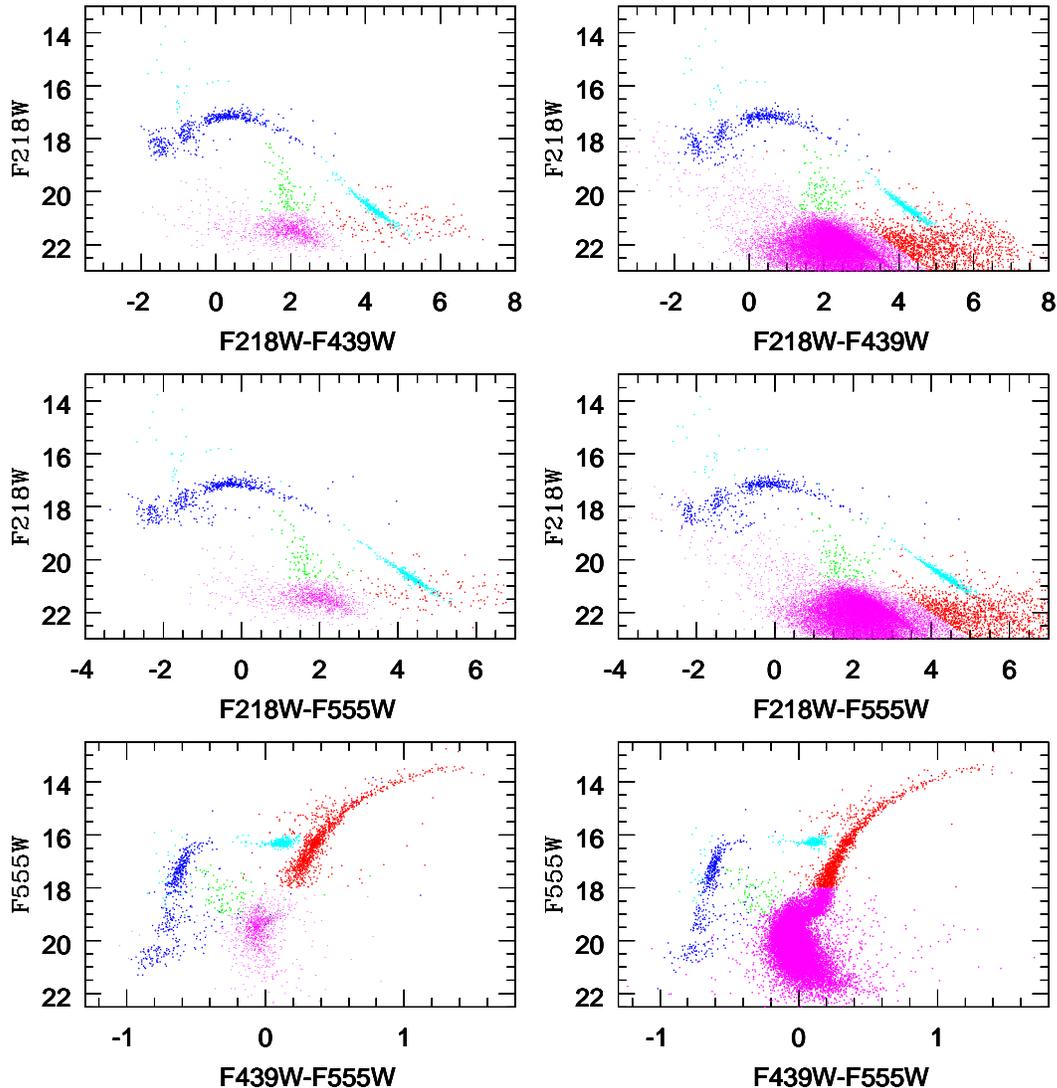}
\caption{Optical-UV Color-Magnitude Diagrams derived using two
different reduction packages, namely ROMAFOT (left), and DAOPHOT
(right) as well as two different reduction strategies concerning
the identification of individual sources. See text for
details.} \label{f:fig1} 
\end{figure*}

As a starting point of the present  investigation we will refer to the
observational  material collected by Bedin et   al (2000), as based on
data   taken with  1.54m  ESO-Danish Telescope  in  the  optical bands
U,B,V,I,  complemented with   similar data   from  the  ESO-NTT
Telescope and with $F439W$ and $F555W$ frames  taken with the WFPC2 on
board of  HST (Piotto et  al.  2002). In   the cluster external region
(r$\ge$ 120 arcsec)  unaffected by the crowding  of stellar images, we
will rely on the Bedin et  al.  (2000) photometry.  Data for the inner
region have  been  reprocessed, with  the  additional use  of  UV data
collected in  the  $F218W$ filter. The  optical and  UV data retrieved
from the HST archive (ID: GO6095, PI:  Sosin) are the following:
$F218W$ (1600, 1700 s),  $F439W$ (50, $2\times230$s), and $F555W$  (7,
50s).

For  data  reduction we  adopted  the   ROMAFOT  package (Buonanno  \&
Iannicola 1989).  Even tough    this package is more  demanding   than
DAOPHOT in terms  of man power, it  allows an interactive check of the
detection and fit accuracy of individual stars in crowded fields.  The
search   was performed on the  $F218W$   data, by adopting a detection
threshold of $5 \sigma$ above  the sky level.   As discussed, e.g.  by
Ferraro et al  (1999), the advantage of using  UV band is that hot  HB
stars appear   as bright objects,  and  the images,  even  in the very
center of the  cluster,  are marginally affected   by the crowding  of
luminous  Red  Giants.  However, this   approach also  means  that the
detection of Red  Giant  stars in  UV data  is  very  challenging.  To
collect a complete sample of  cluster luminous stars, the above search
was complemented with an additional  search on the $F439W$ data,
but only for sources brighter than $F555W\sim$ 17.  The photometry was
eventually performed using  an analytical PSF,  a Moffat function with
$\beta=2$, modeled from a sample of $\sim 30$ isolated stars uniformly
distributed across the field.

As a  test, we repeated  the  reduction with DAOPHOT/ALLFRAME (Stetson
1987, 1994) as explained in Piotto et al. (2002).  The search, in this
case, was performed  on the median of  the $F439W$ and $F555W$ images,
and  then PSF fitting photometry  was  performed simultaneously on all
the frames. The instrumental magnitudes have been transformed into 
the so-called STMAG system of the WFPC2 using the calibration provided 
by Holtzman et al. (1995).  

Figure~1 shows the  CMDs   obtained using  ROMAFOT (left    panels) or
DAOPHOT/ALLFRAME (right panels). As   a consequence of  the  different
reduction  strategies, the number   of faint cool sources ($F218W  \ge
22$, $ F218W-F439W\approx2$) is much  larger in the CMD obtained using
DAOPHOT. However, in the bright region the different CMDs appear quite
similar, with  a clear evidence  for the already known trichotomy of
blue HB stars.

\section{Horizontal Branch structures}

\begin{figure*}
\centering
\includegraphics[width=15cm]{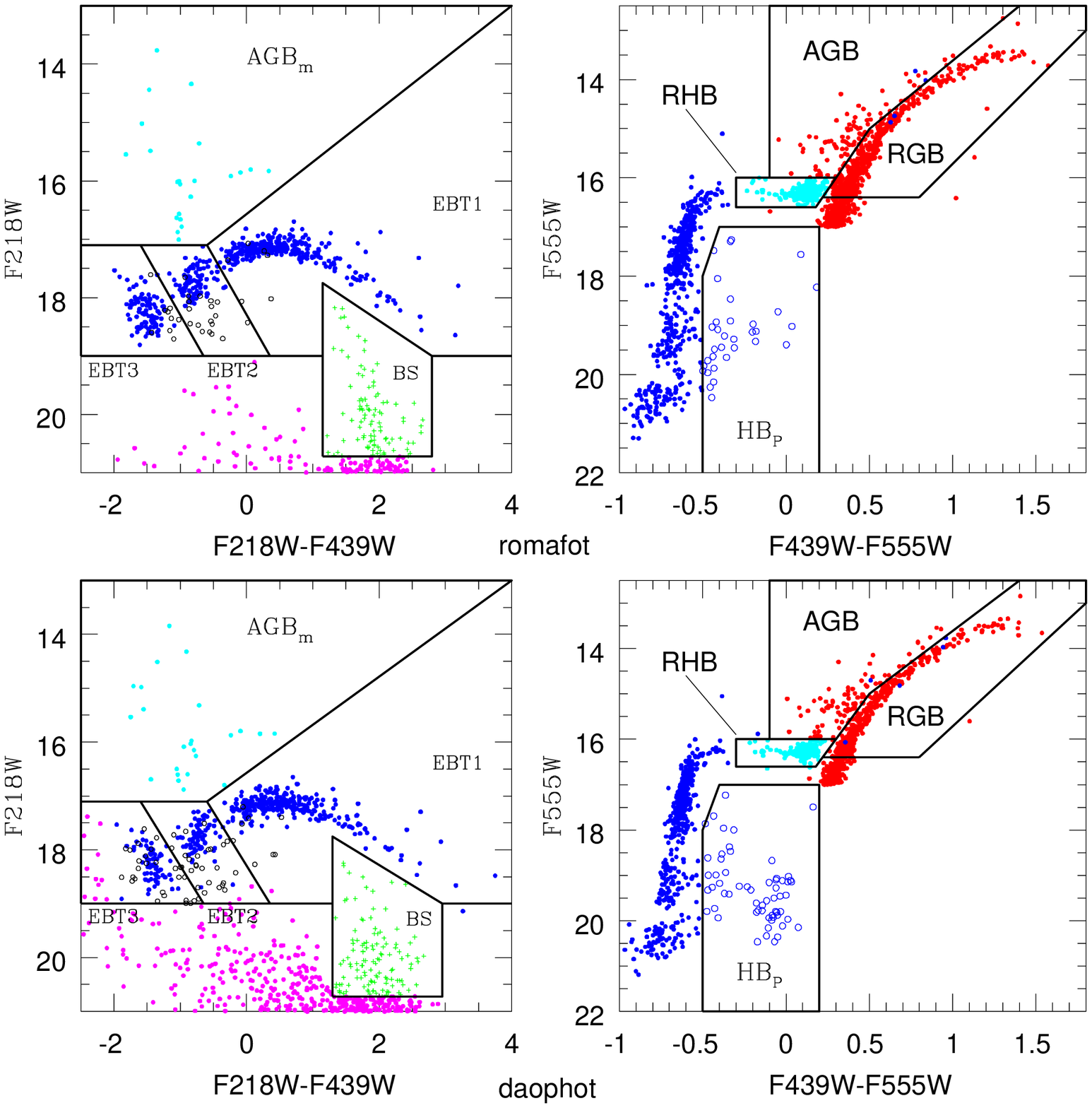}
\caption{Selected regions of the UV (right) and optical (left)
CMDs. Different boxes show magnitude and color ranges adopted to
select the different stellar samples (see labels). Upper and lower
panels display photometric data reduced using ROMAFOT and DAOPHOT, 
respectively. Note that MS, sub-giant-branch, BS, and faint RGB 
stars have been artificially removed to make more clear the selection 
of HB and RGB stars.} \label{f:fig2} 
\end{figure*}

The opportunity to analyse, for the same cluster,  both UV and optical
data allow us to perform a detailed analysis of luminous cluster stars
in the various evolutionary  phases. Fig.~2  shows  the boxes used  to
select the different  samples.  Blue HB  stars, following  Bedin et al
(2000),  have been divided into three   groups, namely EBT1, EBT2, and
EBT3  in order of increasing effective temperature.  
In particular, the $EBT1$ sample includes hot HB stars located between
the blue edge of the RR Lyrae instability strip and the first gap; the $EBT2$
and $EBT3$ samples include extreme HB stars located between the first and
the second gap, and hotter than the second gap; while the $AGB-manque'$ ($AGB_m$)
sample includes low-mass AGB stars brighter than typical HB stars. These
subsamples and Blue Straggler (BS) stars (plus  signs) can be easily
identified in the $F218W$,  $F218W-F439W$ diagram (left panels), since in
this plane they are  brighter and cover a color range of approximately 5  mag.
For the  same  reasons, red  HB (RHB) stars, RGB stars, and  AGB stars  were
selected in  the  $F555W$, $F439W-F555W$ diagram. 
The conclusion of this investigation do not depend on the criteria 
adopted to select individual HB and RGB subsamples. 

However, a glance at the data plotted in Fig.~1 and in Fig.~2 reveals
the  presence of  a   peculiar group of    objects, which in  $F218W$,
$F218W-F439W$ CMD behave either as regular or as subluminous EBT2 or EBT3
stars,  whereas in the  optical  CMD they appear systematically cooler
than typical HB stars. We will call these  peculiar objects HBp stars,
marking their position in the CMD with open circles (see Fig.~2). 
Note that the HBp region shown in Fig. 2 (left panels) would include also
all the BSs population, not shown here. We call HBp, however, only those stars
that lie in the BSS region in optical CMDs, but near the HB in the UV ones
(see right panels of Fig. 2), where the distinction between BSs and HBp is
unambiguous. 
This behavior is  indicative of a composite nature  of the spectrum and can
be  caused either by photometric  blends or by physical binarity (see,
e.g., Allard et  al. 1994).  As a matter  of fact, a detailed check of
individual objects  on optical and  UV images, performed with ROMAFOT,
discloses that the group of cooler HBp ($F555W\approx20.0-20.5$, 
$F439W-F555W\approx-0.2-0.0$) objects detected by DAOPHOT is mainly due 
to blends, i.e. the faint companions located close to these
stars   have  not been detected.  However,   the surviving HBp objects
appearing in the ROMAFOT CMD have  very regular images, and we believe
rather improbable that this could be the  result of casual blends with
foreground  stars.  This hypothesis  is also supported by the evidence
that HBp stars are distributed across the cluster core and do not show
a centrally peaked distribution.  Therefore, HBp  stars might be considered 
as a possible evidence for a binary  nature of at least  a fraction of  EBT
objects.

For reference,  the arrows plotted  in Fig.~3 show the  shift in
the CMD position for a typical  EBT3 or EBT2  stars, caused by a blend
with  a  redder  companion.   To mimic the    most plausible events we
assumed  that the blended  star  (or the companion   star in case of a
physical binary) is a Main  Sequence (MS) star  located just below the
Turn Off,  namely  at $F555W\sim  20$  and $F439W-F555W\sim 0.0$.  Not
surprisingly, one finds a maximum shift in  the faintest EBT3 objects,
a much smaller effect in the EBT2 group and  virtually no shift at all
among stars in the EBT1 group.  The distribution of observational data
convincingly indicates that  the   large majority of  HBp   objects is
indeed   originated from the EBT3 group.    One also concludes that at
least a part of the spread in both EBT2  and EBT3 groups is due either
to binaries or  to blends. In the  hypothesis  of binaries,  we notice
that this affects photometric data only  when the companion of the hot
HB stars is a relative bright H  burning structure. Thus one could not
exclude that a much larger fraction of EBT3 objects might be binaries.
To  constrain  this  working  hypothesis on a  more  quantitative
basis, we performed, several artificial  star experiments. In order to
evaluate the completeness of hot HB stars  we estimated the ridge line
of this sample in the $F218W-F439W, F218W$ plane (see Fig. 1) and then
we  randomly distributed 150 artificial  stars in each WFPC2 chip. The
photometry on  individual images was  performed with ROMAFOT following
the   same reduction     strategy  adopted to    obtain   the original
photometry. We found  that the completeness  is higher  than 99\% over
the   entire magnitude  range ($16.8   \le  F218W  \le 20$).  The same
procedure was adopted to estimate the completeness of HB and RGB stars
brighter than  $F555W\le 17$ in   the $F439W-F555W, F555W$  plane (see
Fig. 1).  The main  difference is that the  number of artificial stars
was increased to 400 per WFPC2 chip. We found that the completeness is
higher than  99\% for $F555W\le  17$ and becomes  equal to 87\% in the
magnitude range $20 \le F555W \le 21$.  According to these findings we
did not   account for   uncertainty due   to completeness,   since the
selection  of  HB  and  RGB subsamples in   different  CMDs presents a
completeness on average  better  that 99\%.   Moreover  and even  more
importantly,  we found that the fraction  of HB stars recovered inside
the HBp  region  of the $F439W-F555W,  F555W$  plane (see Fig.  2), is
smaller than 2\% of the total number of artificial HB stars adopted in
the  experiment (16 vs   1000).   This finding  would imply  that  the
fraction of blends in  the sample of $HB_p$ stars   is at most  of the
order of 50\%.

\begin{figure}
\centering
\includegraphics[width=8cm]{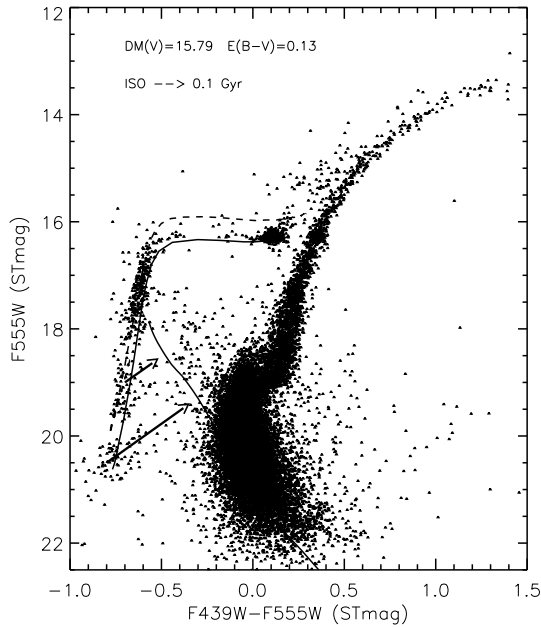}
\caption{The best fit of theoretical predictions to the observed
$F555W$, $F439W-F455W$ CMD. The photometry was performed with DAOPHOT. 
The solid and the dashed line located below and above HB stars display 
the ZAHB and the exhaustion of central He for Z=0.001 and Y=0.23. 
The solid line shows a stellar isochrone for the same chemical composition 
and an age equal to 0.1 Gyr. The arrows show the shift in color and in 
magnitude of different HB stars due to either a companion or a blend 
with a MS star.}\label{f:fig3} 
\end{figure}

\begin{figure}
\centering
\includegraphics[width=8cm]{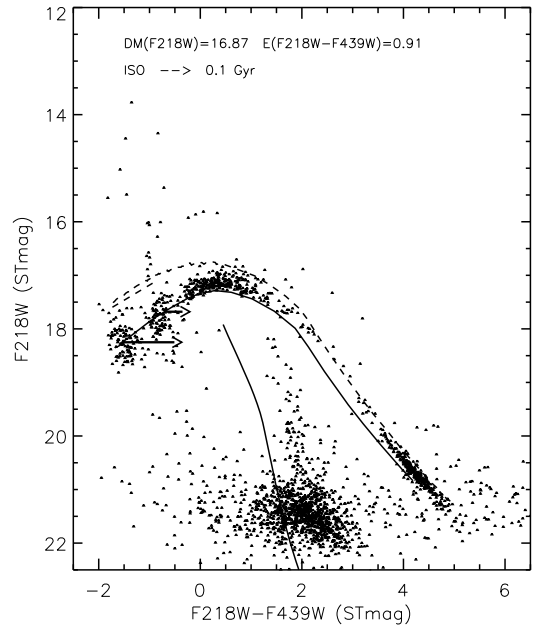}
\caption{Same as in Fig.~\ref{f:fig3} but for the F218W, F218W-F439W 
CMD. The photometry was performed with ROMAPHOT.} \label{f:fig4}
\end{figure}

For comparison with theoretical prediction we  will rely on the recent
set of HB models by   Cassisi et al.  (2004),  by assuming an original
cluster composition of Z=0.001  and Y=0.23 (Harris  1996, but see also
Bedin et al. 2000 for a discussion on the cluster metallicity). 
Figure~\ref{f:fig3}
shows the best fit of the optical CMD as obtained for the distance and
reddening indicated in the labels. We attempted a simultaneous fit of 
the UV CMD (Fig. 4) using the same distance and reddening, and converting 
the E(B-V) to the corresponding extinction in F218W through the reddening 
law given in  Cardelli  Clayton, \& Mathis (1989). However, as already 
found by Sosin et al. (1997), such value did not provide a proper match 
between theory and observations. A reasonable fit to the UV CMD (see 
Figure~\ref{f:fig4}) requires  a small adjustment in reddening, namely 
A$_{F218W}$/E($F439W-F555W$)=3.60, slightly larger than the value given 
by Cardelli et al. (3.23), and a distance modulus $(m-M)_O=15.42$, which 
is only 0.03 mag larger than the one in Fig.~\ref{f:fig3}.

\begin{table*}
\caption{Selected parameters for the  hot end of the theoretical
ZAHB with Z=0.001 and Y=0.23. The mass of the He-core is $M_c= 0.5018 M_{\odot}$.}

\begin{center}
\begin{tabular}{c c c c c c c }
\hline \hline
M & M$_{env}$ & T$_e$ & $F218W$ &$F218W-F439W$ & $F218W-F555W$ & $F439W-F555W$\\
\hline
0.5022 & 0.0004 & 33788 & 1.4847 & -2.448 & -3.348 & -0.899\\
0.5028 & 0.0010 & 31860 & 1.4513 & -2.372 & -3.260 & -0.888\\
0.5040 & 0.0022 & 31160 & 1.4273 & -2.339 & -3.224 & -0.884\\
0.5060 & 0.0042 & 30320 & 1.3881 & -2.301 & -3.180 & -0.879\\
0.5080 & 0.0062 & 29066 & 1.3384 & -2.243 & -3.115 & -0.871\\
0.5100 & 0.0082 & 27742 & 1.2949 & -2.177 & -3.043 & -0.866\\
0.5150 & 0.0132 & 25385 & 1.1947 & -2.046 & -2.896 & -0.850\\
0.5200 & 0.0182 & 23762 & 1.1231 & -1.944 & -2.783 & -0.839\\
0.5250 & 0.0232 & 22469 & 1.0553 & -1.853 & -2.684 & -0.830\\
0.5300 & 0.0282 & 21385 & 0.9951 & -1.770 & -2.593 & -0.823\\
0.5350 & 0.0332 & 20480 & 0.9487 & -1.695 & -2.510 & -0.816\\
0.5400 & 0.0382 & 19542 & 0.8962 & -1.620 & -2.429 & -0.809\\
0.5450 & 0.0432 & 18933 & 0.8572 & -1.551 & -2.354 & -0.803\\
0.5500 & 0.0482 & 18249 & 0.8127 & -1.478 & -2.276 & -0.797\\
0.5600 & 0.0532 & 17021 & 0.7285 & -1.335 & -2.120 & -0.786\\
0.5700 & 0.0632 & 15929 & 0.6506 & -1.188 & -1.963 & -0.775\\
0.5800 & 0.0732 & 14753 & 0.5545 & -1.005 & -1.672 & -0.763\\
\hline \hline
\end{tabular}
\end{center}
\label{t_tab1}
\end{table*}

However, in both Figure 3 and 4  one may appreciate the good agreement
between  observed and theoretical HB  sequences. In particular, Fig.~4
greatly improves the  UV picture of HB  stars in NGC2808 already given
by Sosin et al. (1997), Brown et al. (2001), and Moehler et al. (2004),
since the   present CMD   covers, for  the  first time,   the   entire
temperature range of HB stars,  i.e. from red HB  to the extreme  blue
tail  and with sizable   samples of both  EBT2  and  EBT3 groups.  The
morphology of the hot end of  HB stars has already  been addressed in  
several relevant theoretical  and empirical investigations.  D'Cruz et
al. (1996) first advanced the suggestion for  a connection between the
hotter HB structures  at the predicted  occurrence of  "hot flashers",
namely the stars  which ignite He after  leaving the Red Giant  Branch
due to a huge mass loss. By following Brown et al. (2001, but see also
Cassisi et al.  2003a) the EBT3 group  should be the result  of Helium
flash-induced mixing, enriching the envelopes  of He and C and pushing
the star at effective  temperatures  beyond the  canonical end of  the
Horizontal  Branch   and     at  fainter  magnitudes.    Spectroscopic
investigations by Moehler et al. (2004) have recently supported such a
picture, disclosing that EBT3 stars  are both hotter and more  He-rich
than canonical HB stars,  though revealing an unexpected presence of
atmospheric H.  
According to Unglaub (2004) the occurrence of atmospheric H might 
be due to a mixing of processed material to the stellar surface and to 
the upward  diffusion of H during subsequent evolutionary phases. 

In this context current data raise several questions. According to the
quoted scenario, the canonical ZAHB should stop  at the temperature of
EBT2  stars,  but  Sosin  et al.    (1997)  already  presented a  ZAHB
approaching the effective temperatures typical of EBT3 stars. Brown et
al. (2001) pointed out the problem, and claimed  that this is only the
result of the adopted  ZAHB which reaches  envelope masses as small as
10$^{-4}$  M$_{\odot}$,  i.e. well below  the true  termination of the
canonical ZAHB. According  to Brown et  al. (2001)  the canonical ZAHB
should have a lower limit in the envelope mass equal to M$_{env} \sim$
0.0006 M$_{\odot}$. It is worth mentioning that  such a limit could be
much less firm  than currently assumed,  since Brown et al. (2001) and
D'Cruz et  al. (1996) have explicitly neglected  mass loss in the post
RGB evolution.

However, to further constrain this point we artificially truncated the
canonical ZAHB at an envelope mass that is almost a factor of two larger
than the quoted lower limit ($M_{env}/M_{\odot} \sim$ 0.001 vs 0.0006).
Data plotted in Fig.~4 clearly show that the canonical ZAHB is, against
the expectation, still reaching the EBT3 group. 
To look into  this problem  with some detail, Table
\ref{t_tab1} gives our theoretical predictions for the  hot end of the
ZAHB distribution.  Comparison  of these   values with the    observed
distribution in Fig.~4  reveals  that the blue  boundary of   the EBT1
should lie at T$_e  \sim$ 16.000 K, whereas  EBT2 structures appear in
the  interval 18.000  $\le$ T$_e \le$   23.000 K. In  passing, one may
notice that such a prediction  appears in beautiful agreement with the
spectroscopic temperatures determined   by Moehler et al.  (2004)  for
three  EBT2 structures, namely 20,100,  21,100 and 21,300 K. According
to current HB models, stellar strucutures at the  blue end of the EBT2
region should still have an envelope as large  as M$_{env} approx$ 0.01
M$_{\odot}$, i.e.,  much larger than the quoted  lower limit of 0.0006
M$_{\odot}$. As  a matter of fact, at  this limit our models reach an
effective  temperature  of T$_e \sim$  32.000 K,  well beyond  the red
boundary of the EBT3 group, which lie at T$_e \sim$ 28.000 K.

\begin{table*}
\begin{center}
\caption{Star counts of Red Giant Branch, Horizontal Branch, Asymptotic Giant 
Branch, and Blue Straggler stars according to the photometry performed with 
ROMAFOT.\label{t1}}
\begin{tabular}{ccccccccccccc}
\hline \hline
    r        & RGB & RHB & EBT1&EBT2& EBT3& HBp&HB(tot)& R & $\sigma_R$& AGB&   AGB$_m$ & BS \\
\hline
 HST -- ROMAFOT  \\
0.0 - 15.6   & 163 & 113 &  64 & 22 &  17 &  8 & 224 &1.37 &0.14 & 33 &2 & 43\\
15.6 - 45.6  & 226 & 160 & 119 & 21 &  26 & 16 & 342 &1.51 &0.13 & 30 &8 & 32\\
45.6 - 120   & 199 & 123 & 124 & 50 &  45 & 10 & 352 &1.77 &0.16 & 19 &12& 22\\
\hline
TOTAL &588 & 396 & 307 & 93 &  88 &  34& 918 &1.56 & 0.08  & 82& 22 & 97 \\
\hline
Bedin et al. (1999)    \\
120 - 160    &92  &68 &34 & 23 & 32 &\ldots&157&1.71&0.22 & 16 &\ldots&\ldots\\
160 - 250    &84  &82 &32 & 18 & 32 &\ldots&164&1.95&0.26 & 21 &\ldots&\ldots\\
\hline
TOTAL Bedin  &176 &82 &150& 41 &64  &\ldots&321&1.82&0.17 & 37 &\ldots&\ldots\\
\hline
ROMAFOT+Bedin&764 &478&457&134 &152 &\ldots&1239&1.62&0.07& 119&\ldots&\ldots\\
\hline \hline \hspace*{2.5mm}
\end{tabular}
\end{center}
\label{t:tab2}
\end{table*}

\begin{table*}
\begin{center}
\caption{Star counts of Red Giant Branch, Horizontal Branch, Asymptotic Giant 
Branch, and Blue Straggler stars according to the photometry performed with 
DAOPHOT/ALLFRAME.\label{t1}}
\begin{tabular}{ccccccccccccc}
\hline \hline
  r & RGB & RHB & EBT1&EBT2& EBT3& HBp&HB(tot)& R & $\sigma_R$& AGB&   AGB$_m$ & BS \\
\hline
 HST -- DAOPHOT  \\
0.0 - 15.6   &154  & 110 & 67 & 17 & 16 &  14 & 224 &1.45  &0.15 & 35 & 2 &51\\
15.6 - 45.6  &213  & 158 &127 & 26 & 21 &  20 & 352 &1.65  &0.14 & 31 & 9 &44\\
45.6 - 120   &181  & 124 & 95 & 32 & 35 &  26 & 312 &1.73  &0.16 & 20 &13 &24\\
\hline
TOTAL        &548  & 392 &289 & 75 & 72 &  60 & 888 & 1.62 &0.09 & 86 & 24&119\\
\hline
DAOPHOT+Bedin &724  &474&439   &116 &136 &\ldots&1209&1.67&0.08 & 123&\ldots&\ldots\\ 
\hline \hline \hspace*{2.5mm}
\end{tabular}
\end{center}
\label{t:tab2}
\end{table*}


Moreover, Fig.~4 shows that  a  large fraction  of both EBT2  and  EBT3
groups falls below the  theoretical ZAHB. The underluminosity  of EBT3
structures  has  been   often considered   as  an empirical   evidence
supporting the mixing scenario,  which -however- should not affect the
EBT2 group.  In this paper we have shown  that such an underluminosity
might be partially due   to binarity (or blending).   To reach firm
conclusions one should have a better insight into the role of binarity
in both the location in the CMD and in the spectral features. Last but
not least, the distribution in Fig.~4 appears hardly in agreement with
the predictions of  the mixing scenario,  as suggested by data plotted
in Fig.~4 (panel 'b') of the quoted paper by Brown  et al. (2001).  We
conclude that the problem of hot HB stars in NGC2808 is far from being
settled, and  deserves  further and accurate  investigations. 
It is worth mentioning that current findings support the alternative 
hypothesis, originally suggested by Bailyn \&  Iben (1989), that cluster 
sdB stars might be the aftermath of mergers of low-mass helium white dwarfs 
formed from primordial main sequence binaries.   

One may finally note that the Blue Straggler sequence ($F218W \le 21$,
$1.5 \le F218W-F439W \le 2.5$) follows the predicted 0.1 Gyr isochrone,
but clearly  shifted toward redder colors. This occurrence can also be
detected in the STIS UV CMD provided by Brown et al. (2001, see their
Fig. 3). For a detailed discussion concerning the evolutionary properties
of cluster BSs the reader is referred to Piotto et al. (2004). 

\section{Star counts and Population Gradients}

To   properly investigate the  radial  distribution  of the  different
samples we split the field covered by WFPC2 in concentric annuli.  The
annuli were  selected  at  the core    radius (15.6 arcsec),   at  the
half-mass radius (45.6 arcsec), while the third one to the edge of the
WF chip, namely  120  arcsec. Structural parameters for  NGC~2808 come
from the compilation by  Harris (1996).  Table \ref{t:tab2} gives star
counts  for the   various selected  phases and  in  the three selected
annuli.  In  the same table current  data  were complemented with star
counts based on  ground-based  U, and optical   B,V data collected  by
Bedin et al.  (2000). Note that  to avoid spurious contaminations from
field stars  we restricted the  star  counts to two  annuli that range
from 2 to  roughly 4 arcmin from   the cluster center (see  Fig.~16 in
Bedin et al. 2000). As a result, one finds  that we are dealing with a
sample of more than 1200 HB stars, i.e. one of the most statistically
significant sample of HB stars in GGCs.

\begin{figure}
\centering
\includegraphics[width=8cm]{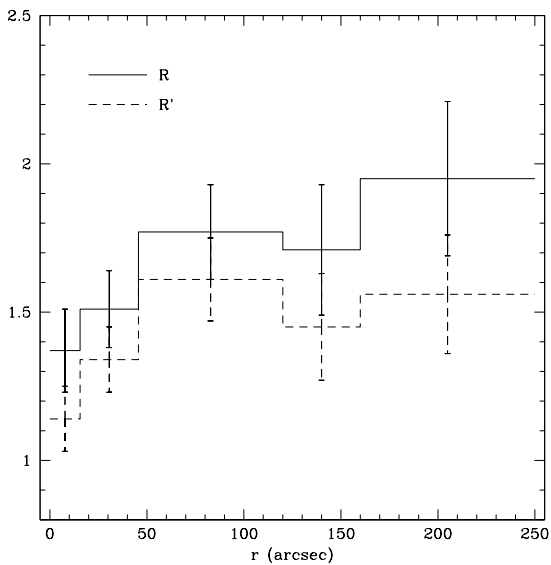}
\caption{The parameters R  (solid line) and $R'$ (dashed line) as a function 
of the radial distance according to ROMAFOT photometry.}
\label{f:fig5}
\end{figure}

To figure  out whether  the evolved  population  presents any peculiar
trend    we    estimated   the R  and      the $R'$  parameters,  i.e.
$R=N_{HB}/N_{RGB}$,   $R'=N_{HB}/(N_{RGB}+N_{AGB})$ where the sample of
cool giants  refers to structure more luminous  that the HB luminosity
level. The  HB luminosity was fixed 2  $\sigma$ below the peak  in the
luminosity distribution of red HB stars, i.e. F555W=16.4 mag. Figure 5
shows the radial  distribution of R. Data  plotted in this figure show
that  these parameters steadly  increase  when  moving from the   very
center  to the outermost  regions, passing  from  $1.37\pm0.14$ in the
cluster   core,  to   $1.51\pm0.13$   at half-mass  radius,    and  to
$1.95\pm0.26$ in the outer regions. It  is worth mentioning, that this
finding is indipendently  supported  by  star   counts based  on   the
DAOPHOT/ALLFRAME cluster photometry (see data listed in Table 3).

Note that in the outermost regions, ground-based photometry is missing
the  HBp    group, thus underestimating    the  actual  number  of  HB
structures.  This means   that the two   outermost R  values should be
considered   as    (marginally) underestimated.  Figure~5    shows the
presence of a population gradient, for  which the number ratio between
HB  and RGB stars   increases when moving from  the  center toward the
cluster periphery.  
This finding supports the color gradient detected by Sohn et al. (1998).
In particular, they found that the cumulative color distribution ranges
from $U-B$=0.4 in the very center of the cluster to $U-B$=0.2 at
$r \ge 70$ arcsec (see their Fig. 1). We already mentioned that both
Walker (1999) and Bedin et al. (2000), did not detect a well-defined population
gradient. However, Walker by using accurate ground-based data found that the
ratio of AGB plus RGB stars brighter than V=15 with the RGB stars with
$15 \le V \le 17.5$ ranges from 8.8 for $r \ge 65$ arcsec, to 5.4 for
$22 < r <65$ arcsec, and to 1.4 for $r\le 22$ arcmin. No firm conclusion was
reached due to the crowding in the innermost cluster regions. 
Moreover, Bedin et al. by using both ground based and HST data did not detect
a statistically significant population gradient, but a possible decrease of RGB
stars for $r\le 100$ arcsec, and a lack of EBT3 stars for $r\approx 250$ arcsec
(see their Fig. 17).
The main difference between the present analysis and the one by Bedin et al. is that
we performed the initial search for hot HB stars in the $F218W$-band images,
hence our completeness is higher in this filter. 

\begin{figure}
\centering
\includegraphics[width=8cm]{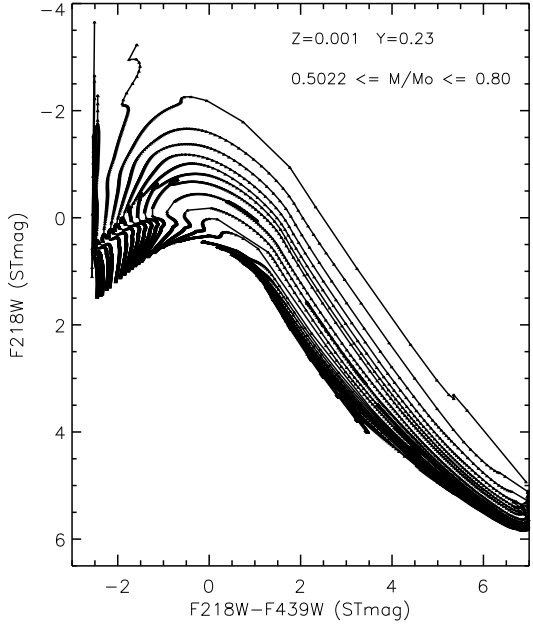}
\caption{The predicted evolutionary tracks from the ZAHB of the
adopted stellar models. }
\label{f:fig6}
\end{figure}

The evidence for very hot HB structures, approaching the limiting mass
for He ignition,  can be taken as  an evidence that some cluster stars
are  expected   to follow  different    types of peculiar evolutionary
histories. A  first  and well-known occurrence   is that very  hot  HB
structure  will miss their   proper  AGB phase,   crossing the CMD  as
luminous    "AGB-manqu\`e"  structures    to   reach at  even   larger
temperatures   their cooling   radius   as  Carbon-Oxygen  (CO) core 
white dwarfs (WDs). However, if  mass loss  has brought some   Red Giants at 
the limiting mass for  He ignition, it appears quite difficult that the 
same mechanism has also  produced Red Giant stars below that  limit. As a 
consequence, one foresees the additional occurrence of RG missing their 
HB  progenies, directly cooling down as He-core WDs. We note that  the cooling
times of He-core WDs are longer than that of CO-core WDs\footnote{The 
lifetime for a typical CO-core WD with $M=0.5 M_\odot$, and $M_B \sim 11.5$ 
is equal to $t\approx 3.5\times10^8$ Myr, while for a He-core WD with 
$M=0.3 M_\odot$, and $M_B \sim 11.5$ is roughly a factor of two longer
($t\approx 6.7\times10^8$ Myr).}. This intrinsic
difference might be the crucial  parameter to identify the contribution
of similar structures in NGC2808  or in  other clusters with  extended
Blue Tails.

Bearing in mind such a  scenario, one can  use star counts to test the
level of   agreement   between  theory  and    observations. From  the
theoretical side,  Fig.~6 summarizes the evolutionary  fates of the HB
structures we are dealing  with. From  evolutionary times one  derives
that for HB stars  producing AGB progenies  one expects a number ratio
N(AGB)/N(HB)$\sim$0.11 (see, e.g., Cassisi  et  al. 2004). The  number
ratio between AGB-manqu\`e and their  hot HB progenitors should again
be of  the  order of 0.10. As  far  as the  empirical  measurements is
concerned, one finds  that both RHB and  EBT1 groups should enter  the
AGB phase,  whereas EBT2, EBT3  and HBp  should  produce AGB-manqu\`e
structures before cooling down as WDs. A glance at  the data listed in
Table       1   shows   that     N(AGB)/N(HB)=0.116    $\pm$0.01   and
N(AGB$_m$)/N(HB)=0.102$\pm$0.02, and therefore  in good agreement with
current predictions.

Ome may finally notice that the number ratio between BS and HB stars is 
of the order of 0.1, thus confirming the relative lack of BSs found 
by Piotto et al. (2004) in clusters with total luminosity larger than 
$M_V \le$ -9.

\section{HB gradients, Blue Tails and the second parameter: a discussion}

The evidence for the possible occurrence of binary systems at least in
the EBT3  group  appears to us  as  the most relevant outcome  of this
investigation. It suggests that binarity could be at the origin of the
huge mass   loss  which removed their RG  progenitors  of a substantial  
fraction of their H rich  envelope. It is true that in the present photometry 
only a few extreme HB stars show the fingerprint of binarity, but note that MS 
companions significantly fainter than TO stars would not produce significant  
shifts in the magnitude of HB stars. Therefore, extremely hot cluster HB
stars behave like  their Galactic field  counterparts, the sdB  stars,
where  the binarity plays a  relevant role (see Han   et al. 2003, and
references therein).   In this   context  one could  speculate  that a
possible  scenario accounting for  the peculiar trichotomy  of the blue
HB. Stars in the EBT1 group come from a mass  loss mechanism acting on
single stars. Hotter EBT2 and EBT3 stars, or at least a fraction of them, 
have suffered larger mass loss, possibly due to mass exchange with a 
companion that pushes these stars toward the hot end of the HB. 
However, in  the case  of  extreme mass loss,  stars will ignite He 
as "hot flashers"  spending their  HB   lifetime as a separate  group  
of stars  (EBT3) at  the  hot end   of the  normal HB (Sweigart 1997; 
Momany et al. 2004).

As far  as  the occurrence  of radial  gradients is  concerned, it has
already been found an  expanded spatial distribution  of EHB in  a few
GGCs, such as $\omega$ Cen (D'Cruz et al.   2000) and NGC6101 (Marconi
et al. 2001). If one assumes that the gradient of R in NGC2808 is real,
one can further speculate about the  possible origin of blue HB stars,
at least to fix the terms for future developments.   As a first point,
since Iben (1968) the R parameter has been widely used as an indicator
of the original  He content  of HB progenitors.  Even though  the
theoretical calibration  of R in terms  of the original helium  is far
from being firmly settled (see, e.g.  Cassisi  et al. 2003b), the most
straightforward interpretation of the R gradient  would be in terms of
He: the original He content of HB stars increasing  from the center to
the cluster  periphery. However,  it  is hard to   believe to an   
original spatial gradient in the He abundance and to its survival over
the many Gyrs of the cluster lifetime. Therefore we believe that the He
gradient   hypothesis has to  be  ruled out.  Similarly, the dynamical
relaxation of the less massive HB  stars cannot be invoked, because it
has been repeatedly shown that the HB lifetimes are not long enough.

Bearing in mind that HB  stars are the progenies of  RG stars, the two
populations must have  the same distribution,  unless something happen
at  the passage form  RG to HB structures.  As  odd as  it may be, one
should conclude that something has produced a huge amount of mass loss
in some RG stars, at the  same time ejecting these progenitors outside
the cluster core.  This could  be  a continuous  process, refurbishing
continuously the  cluster halo with BHB.   However, one  could also be
tempted to connect   such a mechanism   with a  past  episode  of core
collapse, when  the    tidal stripping of  RG  may    be suddenly  and
enormously enhanced. In the mean time, the new born hot HB stars would
be rapidly thermalized, the actual spatial distribution being possibly
a signature of this dynamical event. In this sense, we do not give for
sure  that the expanded  distribution   necessarily argues against   a
dynamical origin, as claimed by D'Cruz et al. (2000).

As recently pointed out by Rosenberg et  al (2004) discussing the case
of   M54, the most striking   and undeniable  evidence concerning Blue
Tails,   is that they   appear in globulars  which are  among the most
massive   clusters     in  our   Galaxy    (see    also    Moehler  et
al. 2004).  However,  not all   massive clusters  have  Blue Tails. If
cluster  dynamics plays a significant role,   one should conclude that
Blue  Tails  are a  transient    phenomenon rather than a   continuous
one. The occurrence  of this  phenomenon might  be  connected with the
already mentioned  core   collapse, a  dynamical  phase  during  which
stellar  interactions  and tidal stripping   surely achieve a enormous
efficiency. Against such  hypothesis,  one could invoke  the  evidence
that  NGC2808   does      not  show  the   typical     signature    of
post-core-collapsed  clusters (Djorgovski et al.   1991; Fusi Pecci et
al. 1993). However, Meylan \& Heggie  (1997) have already advised that
the   separation   between "King  model clusters" and "core-collapsed
clusters" may often be less clear-cut than generally  believed, since 
we do not exactly know how a cluster appears between two episodes
of collapse. Moreover, one may put the problem on very general grounds
recalling, again    from Meylan \&  Heggie (1997),    "the outstanding
problem implied by the many clusters -like 47 Tuc- which show no trace
of core-collapse, but have short enough dynamical times for collapsing  
in a short fraction of the Hubble time".

According to such a picture, we suggest that Blue Tails and the second
parameter problem could be quite different evolutionary features. The 
former connected with the dynamical interaction in massive clusters, 
whereas the latter possibly connected with an age difference as recently 
dicussed by Rey et al. (2001) for M3 and M13.

\section{Final remarks}

We  have already quoted the uncertainties affecting the theoretical
calibration  of  the  R parameter in  terms of   the original  helium
content. It is  worth  noticing, the  accompanying  uncertainty on  the
empirical estimates. As a matter of  fact, recent investigations focussed
on this problem indicate for NGC2808 an R parameter equal to 0.99 $\pm$
0.1 (Sandquist 2000) or R=1.26 $\pm$ 0.06 (Zoccali et al. 2000) against
the value  R=1.62 ($\pm 0.07$) derived in this paper. It appears that
current  estimates of statistical errors   for the R  parameter are of
minor relevance when compared with completeness errors of the sample, 
at least when dealing with clusters with extended Blue Tails.

Bearing in mind the  caveat  concerning theoretical calibrations,  one
may finally notice the good agreement of present R-value with the most
recent calibration given by Cassisi,  Salaris  \& Irwin (2003),  which
for      [M/H]$\sim$    -1.03   predits     R$\sim$1.55  (Zoccali   et
al. 2000). However, strictly speaking that calibration is adequate for
clusters with  HB structures not  hotter than the RR Lyrae instability
strip, since  evolutionary times  of  very blue HB structures  can  be
longer up to  about 40\% (Cassisi  et al. 2004).  However, according to
the  evidence that EBT2 and  EBT3 stars are  approximately the 30\% of
the entire HB sample, one  easily derives that the  value of the above
quoted  calibration should be   increased by no  more than  19\%, thus
improving the agreement between theory  and observations. This, in our
opinion, can  be taken  as an independent   support to  the $^{12}C  +
\alpha$ reaction rate adopted in the theoretical calibration.

\section{Acknowledgments}

It is a pleasure to thank C.E. Corsi, I. Ferraro, and L. Pulone for several 
useful suggestions concerning the artificial star experiments. We also 
acknowledge Dr. Ivo Saviane as referee for his pertinent comments that 
helped us to improve the content and the readability of the manuscript.
This work was partially supported by MIUR-COFIN~2002 under the project 
"Stellar Populations in Local Group Galaxies" and MIUR-COFIN~2003 under 
the project "Continuity and Discontinuity in the Galaxy Formation". 
MZ thanks Giampaolo Piotto for providing financial support and computer 
facilities when part of this investigation was carried on.


\end{document}